\documentclass[aps,prl,groupedaddress]{revtex4-1}

\usepackage{graphicx}
\usepackage{dcolumn}
\usepackage{bm}

\begin{document}


\title{Peculiar Behavior of Si Cluster Ions in Solid Al}



\author{S. Kawata}
\email[]{kwt@cc.utsunomiya-u.ac.jp}
\affiliation{Graduate School of Engineering, Utsunomiya University, Japan}
\author{C. Deutsch}
\affiliation{Laboratoire de physique des gaz et des plasmas (LPGP), Universit{\'{e}} Paris-Saclay, France}
\author{Y. J. Gu}
\affiliation{ELI Beamlines, Prague, Czech Republic}



\date{\today}

\begin{abstract}
A peculiar ion behavior is found in a Si cluster, moving with a speed of $\sim0.22c$($c$: speed of light) in a solid Al plasma: the Si ion, moving behind the forward moving Si ion closely in a several \r{A} distance in the cluster, feels the wake field generated by the forward Si. The interaction potential on the rear Si may balance the deceleration backward force by itself with the acceleration forward force by the forward Si in the longitudinal moving direction. The forward Si would be decelerated normally. However, the deceleration of the rear Si, moving behind closely, would be reduced significantly, and the rear Si may catch up and overtake the forward moving Si in the cluster during the Si cluster interaction with the high-density Al plasma. 

\end{abstract}

\pacs{36.40.Gk, 36.40.Wa, 52.25.Mq}

\maketitle


The cluster ion interaction with matter has attracted serious attentions in the scientific areas, including a cluster interaction with intense lasers to induce the Coulomb explosion \cite{Wang2000, Wang2018, Ditmire1, Ditmire2}, fast clusters' interaction with targets based on the recent accelerator technology advances \cite{Burunelle1, Takayama1}, and a cluster ion beam application to inertial fusion driver. In ion beam inertial fusion application especially, a preferable ion speed may be about $0.1c\sim0.25c$, that is, around $10\sim50$Mev/u \cite{Kawata1, Clauser1, IHofmann1}. Instead of ion beam, the cluster ion beam inertial fusion has been also proposed \cite{Tahir1, Deutsch1}. In the cluster ion beam (CIB) driven inertial fusion (CIF), it is expected to deposit the cluster beam energy in a small volume of the energy absorber of an inertial fusion fuel pellet by the correlated ion stopping enhancement in plasmas \cite{Zwicknagel1, Bret1, Nardi1, Nardi2, Ben-Hamu1, Arista1, Brandt1}. 

In the correlated cluster stopping power of a plasma, the Coulomb explosion of the cluster, the vicinage effect on the cluster stopping and the wake field by the fast cluster have been studied intensively \cite{Wang2000, Wang2018, Ditmire1, Ditmire2, Zwicknagel1, Bret1, Nardi1, Nardi2, Ben-Hamu1, Arista1, Brandt1}. If the clusters deposit their energy in the material more than the individual ions, the ratio $q/A$ of the charge $q$ and the mass $A$ can be reduced in CIF, and the CIB space charge effect on the beam transport may be also reduced inside the accelerator and during the CIB transport  \cite{Horioka1,Takayama1}. 

In the context of CIF, CIBs interact with a solid material of Al or Au or so, depending on the fuel target design \cite{Kawata1, Horioka1, Tahir1, Deutsch1}. In this paper Si clusters and the solid Al target are employed to study the detail behavior of each cluster ion in the high-density plasma. The typical Si ion distance $l_{c}$ of Si clusters is about 5\r{A}$=5\times10^{-10}$m. In the high-density sold metal Al, the Si cluster ion charge is quickly shielded by the Al free electrons. The solid Al electron number density is about $1.79 \times 10^{29}/m^{3}$, and the Debye shielding length $\lambda_{De}$ is typically smaller than $l_{c}$. After the significant heating by the intense CIBs in CIF, the target temperature may become a few hundreds eV \cite{Kawata1, Tahir1, Deutsch1}. Even at the electron high temperature of 100eV, $\lambda_{De} < l_{c}$. In this paper, we found that the vicinage effect is limited to the couple Si ions moving in parallel to the moving direction among Si cluster ions in the solid Al, and the wake field, generated by one Si ion, influences the vicinage Si ion, just behind the Si ion involved. During the propagation in the solid Al the Si ion behind the forward moving Si closely catches up and overtakes the forward moving Si ion. This interesting behavior of each Si ion in one Cluster is found and presented in this paper. 

On the other hand, the fast single-ion behavior in the high-density plasma has been also studied intensively \cite{Chenevier1, CLWang1, Echenique1, Ichimaru1}. In a dense plasma, assume one fast-moving ion. First the total electric potential, generated by a swift ion $q\delta(\vec{r}-\vec{v}_{0}t)$, is estimated: $\phi_{total}=\phi_{ind}+\phi_{ext}$. Here $\vec{v}_{0}$ is the velocity of the ion. In our case, $v_0/\sqrt{T_{e}/m_{e}}>>1$, and typically $v_0/\sqrt{T_{e}/m_{e}}=\sim5-15$. Here $T_e$ is the electron temperature and $m_e$ the electron mass. In the linear response framework \cite{Ichimaru1}, we can easily describe $\phi_{total}$ in the Fourier space. 

\begin{equation}
\phi_{total}(\vec{r}, t)=\frac{q}{(2\pi)^{3}4\pi\varepsilon_{0}}\int_{-\infty}^{\infty} d^{3}k \frac{1}{k^2}\frac{1}{\epsilon(\vec{k}, \vec{k}\cdot\vec{v}_{0})}e^{i\vec{k}\cdot(\vec{r}-\vec{v}_{0}t)}
\label{eq1}
\end{equation}
\\*
Here $\epsilon(\vec{k}, \vec{k}\cdot\vec{v}_{0})$ is the linear response function of the target plasma electron. The response function would be described as follows: 

\begin{equation}
\epsilon(\vec{k}, \vec{k}\cdot\vec{v}_{0})=1+\left(\frac{k_{De}}{k}\right)^2W\left(\frac{\vec{k}\cdot\vec{v}_{0}}{k\sqrt{T_{e}/m_{e}}}\right)
\label{eq2}
\end{equation}
\\*
Here $W(z)$ is the $W$ function or the Voight function, and $W(z)=\exp(-z^2)erfc(-iz)$. Here we assume $T_{e}\gg E_{F}(=\frac{1}{2}mv_{F}^2)$, where $E_{F}$ is the Fermi energy and $v_{F}$ the Fermi velocity. From $\phi_{total}(\vec{r}, t)$,the induced potential is obtained: $\phi_{ind}(\vec{r})=\phi_{total}(\vec{r})-\phi_{ext}(\vec{r})$. At $\vec{r}=\vec{r}_{0}+\vec{v}_{0}t$, the force acting on the swift ion is obtained by the spatial derivative of $\phi_{ind}(\vec{r})$. Then the stopping power is also obtained in the material. 

For the rest ion of $\vec{v}_{0}=0$, $W\left((\vec{k}\cdot\vec{v}_{0})/(k\sqrt{T_{e}/m_{e}}\right)=1.0$, and $\phi_{ind}(\vec{r}, 0)$ becomes the static Debye screened potential of $\frac{q}{4\pi\varepsilon_{0}}\exp\left(-\frac{r}{\lambda_{De}}\right)$, which is isotropic. When ${v_{0}}/{\sqrt{T_{e}/m_{e}}}>>1$, Eqs. (\ref{eq1}) and (\ref{eq2}) may show that the screening length depends on the spatial direction \cite{Chenevier1,Ichimaru1}. Here we assume $\vec{v}_{0} \parallel x$. For $X=x-v_{0}t>0$, the total electric potential becomes almost the bare Coulomb potential \cite{Chenevier1, CLWang1}:  $\phi_{total}(x)=\frac{q}{4\pi\varepsilon_{0}}\frac{1}{x}$. For $X=x-v_{0}t<0$, $\phi_{ind}(r=0, X)=\phi_{total}(0, X)-\phi_{ext}(0,X)\propto \sin\frac{(x-v_{0}t)/\lambda_{De}}{{v_{0}}/{\sqrt{T_{e}/m_{e}}}}$ \cite{Chenevier1}. The wake field, generated by the swift ion, extends in the $-X$ direction wider than $\lambda_{De}$, depending on ${v_{0}}/{\sqrt{T_{e}/m_{e}}}$. In the transverse direction at $\vec{k}\cdot\vec{v}_{0}\sim 0$, $W\left(\frac{\vec{k}\cdot\vec{v}_{0}}{k\sqrt{T_{e}/m_{e}}}\right)\sim1$, and the usual Debye screening is reproduced. Therefore, the total electric potential $\phi_{total}$, produced by the swift ion $\left({v_{0}}/{\sqrt{T_{e}/m_{e}}}>>1\right)$, is anisotropic and is elongated in $-X=-(x-v_{0}t)$ in our case. The wake field may also influence each ion motion of a cluster and on the stopping power for a cluster in a material. 

In this discussion we assumed non-degenerated electrons. In the context of the inertial fusion \cite{Kawata1, Clauser1, IHofmann1, Tahir1, Deutsch1}, the ion beam or cluster ion beam input total energy is the order of MJ, and the input pulse length is the order of $\sim$10ns. Therefore, the temperature of the energy-absorber material of the fuel target becomes quickly high, for example, a few hundreds eV: $T_{e} \gg E_{F}$ \cite{Kawata1, Clauser1, Tahir1, Deutsch1}. The discussions above would be valid during the interaction time of the body of the input ion beam, except the very initial period. At the initial time the target energy absorber material, say Al or Pb or so would be degenerated. In that case, the electron temperature $T_{F}$ should be introduced in stead of $T_{e}$, and also the response electron number should be reduced along with the Fermi degeneracy \cite{Ichimaru1, Arista2, Maynard1, Arista3}. For example, at the room temperature of $T_{e}\sim 300K$ the number density of the electrons contributing to the interaction with the cluster is reduced by about $10^{-6}$ \cite{Kittel1}. In this specific situation, the relation of $\lambda_{De} \gg l_{c}$ would be fulfilled, and the perfect vicinage effect would appear in the cluster interaction with the solid metal. In inertial fusion target implosion the degenerated phase appears only at the very initial instant.  

\begin{figure}
 \includegraphics[width=53mm, angle=0]{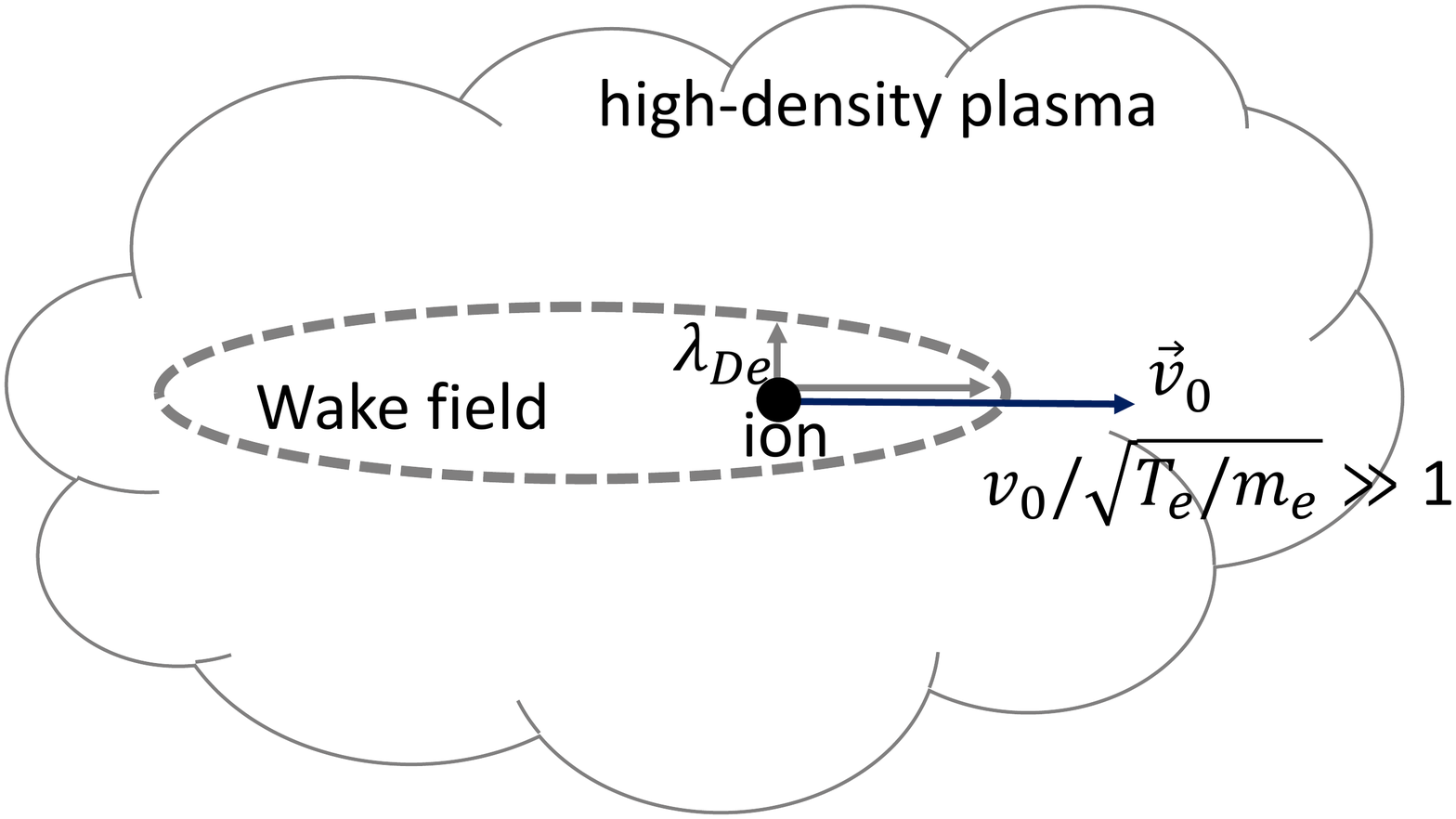}%
 \caption{Schematic diagram of the electric field. In a high-density solid Al, the wake field is localized in transverse and elongated behind the Si ion.  \label{fig:Fig1}}
\end{figure}

\begin{figure}
 \includegraphics[width=80mm, angle=0]{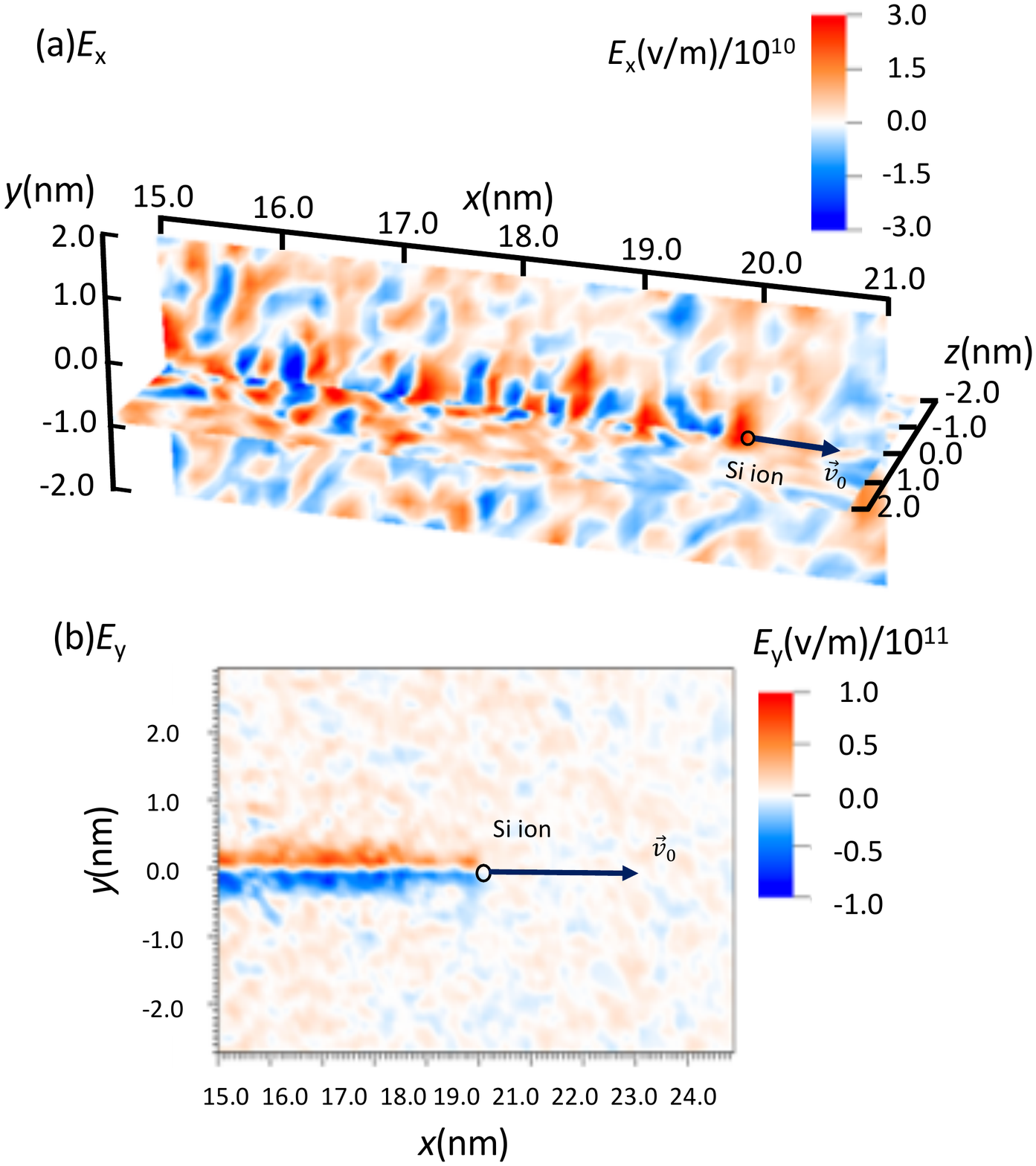}%
 \caption{(a) $E_{x}$ and (b) $E_{y}$ generated by a single Si ion moving in a solid Al with a speed ($v_{x}$) of $0.221c$.  \label{fig:Fig2}}
\end{figure}

Figure \ref{fig:Fig1} shows a schematic diagram of the electric field structure, and Figs. \ref{fig:Fig2}(a) and (b) present the electric field $E_{x}$ and (b) $E_{y}$ generated by a single Si ion moving in a solid Al with a speed ($v_{x}$) of $0.221c$. In Figs. \ref{fig:Fig2}(a) and (b), we computed the electric field $E_{x}$ and $E_{y}$ by using a PIC (particle-in-cell) code of EPOCH \cite{EPOCH1, EPOCH2}. In the PIC simulations the electron behavior is not treated by the dielectric response function, but each electron motion is described by the relativistic equation of motion. In this specific case the Al ion density is solid and the electron temperature is set to be 10eV. The ionization degree of Al is 3 in this case. As shown above, the electric field is confined around the Si ion in the forward direction and also in the transverse direction. In the rear side of Si, that is,  $Z=z-v_{0}t<0$, the wake field appears behind Si with a long tail. In this case, the Debye shielding length is $\lambda_{De}\sim 1\r{A}$. 

The simulation results confirm the analytical estimation results shown above: the electric field is confined around the Si ion in the forward and transverse directions in the distance smaller than the Si cluster ion distance $l_{c}\sim 5\r{A}$. However, the tail of the wake field extends beyond $l_{c}$. Therefore, The vicinage effect of a Si cluster is expected in the longitudinal direction ($\parallel \vec{v}_{0}$) in the solid Al.  

The vicinage effect on a cluster appears \cite{Wang2000, Wang2018, Zwicknagel1, Bret1, Nardi1, Nardi2, Ben-Hamu1, Arista1, Brandt1}, when all or a part of ions consisting the cluster are located inside of the electric field area generated by the cluster ions; the size of the wake field area shown by a dotted ellipse in Fig. \ref{fig:Fig1} should be larger than the cluster size so that the cluster ions behave as one accreted huge ion. 

Here we start to think about a cluster interaction with a dense plasma. The stopping power  would be described by $-{dw}/{dx}=-\sum_{j}\left({\vec{v}_{j}}/{v_{j}}\right)\cdot\vec{F}_{j}$.
The summation is carried out over all the ions consisted of the cluster. Here $w$ shows the total cluster ion energy, and $\vec{F}_{j}$ the force acting on each ion consisting of the cluster. The total cluster charge density $\rho$ is $\rho = \sum_{j}\rho_{j}=\sum_{j}q_{j}\delta(\vec{r}-\vec{r}_{j0}-\vec{v}_{j}t)$. Here $\vec{r}_{j0}$ is the initial position of the $j$th ion of the cluster. Here $\vec{F}_{j}=q_{j}\vec{E}(\vec{r}_{j})
=q_{j}\sum_{i}\frac{q_{i}}{(2\pi)^{3}\varepsilon_{0}}\int d^{3}k \frac{\vec{k}}{k^{2}}\mathrm{Im}\left[ \frac{1}{\epsilon(\vec{k}, \vec{k}\cdot\vec{v}_{i})}\right] \cos(\vec{k}\cdot\vec{r}_{ij}).$

 \begin{eqnarray}
-\frac{dw}{dx}&&= \sum_{ji}\frac{q_{j}q_{i}}{(2\pi)^{3}\varepsilon_{0}}\cdot \nonumber\\
&&\int d^{3}k \frac{\vec{k}\cdot\vec{v}_{j}}{k^{2}v_{j}}\mathrm{Im}\left[ \frac{1}{\epsilon(\vec{k}, \vec{k}\cdot\vec{v}_{i})}\right] \cos(\vec{k}\cdot\vec{r}_{ij}).
\label{eq3}
\end{eqnarray}

Here $\vec{r}_{ji}=\vec{r}_{j}-\vec{r}_{i}$. If we can assume $\vec{v}_{j}\sim\vec{v}_{i}\sim\vec{v}_{0}$ (the averaged cluster ion speed), Eq. (\ref{eq3}) is simplified to be the followings: 

 \begin{eqnarray}
&&-\frac{dw}{dx}= \frac{1}{(2\pi)^{3}\varepsilon_{0}}\cdot \nonumber\\
&&\int d^{3}k \frac{\vec{k}\cdot\vec{v}_{0}}{k^{2}v_{0}}\mathrm{Im}\left[ \frac{1}{\epsilon(\vec{k}, \vec{k}\cdot\vec{v}_{0})}\right] \sum_{ji}q_{j}q_{i}\cos(\vec{k}\cdot \vec{r}_{ji}). 
\label{eq4}
\end{eqnarray}

In Eq. (\ref{eq4}), $\sum_{ji}q_{j}q_{i}\cos(\vec{k}\cdot\vec{r}_{ji})=\sum_{i}q_{i}^{2}+\sum_{j \neq i}q_{j}q_{i}\cos (\vec{k} \cdot \vec{r}_{ji})$. The second term of the right-hand side in the expression shows the correlation between two of the cluster ions. In the previous studies of the vicinage effect on the stopping power for the cluster, frequently this approximation is employed \cite{Wang2000, Wang2018, Zwicknagel1, Bret1, Nardi1, Nardi2, Ben-Hamu1, Arista1, Brandt1}. For the perfect vicinage effect, the stopping power enhancement factor is obtained as $\left( N+{}_N C _2\right)/N=\left(N+N(N-1)\right)/N=N$, where $N$ shows the total ion number in one cluster, which consists of identical ions. In our case this approximation is not valid, because the solid Al electron number density is so high that the wake field generated by a Si ion does not cover all or most of other cluster ions in one cluster, as discussed above. 

\begin{figure}
 \includegraphics[width=90mm, angle=0]{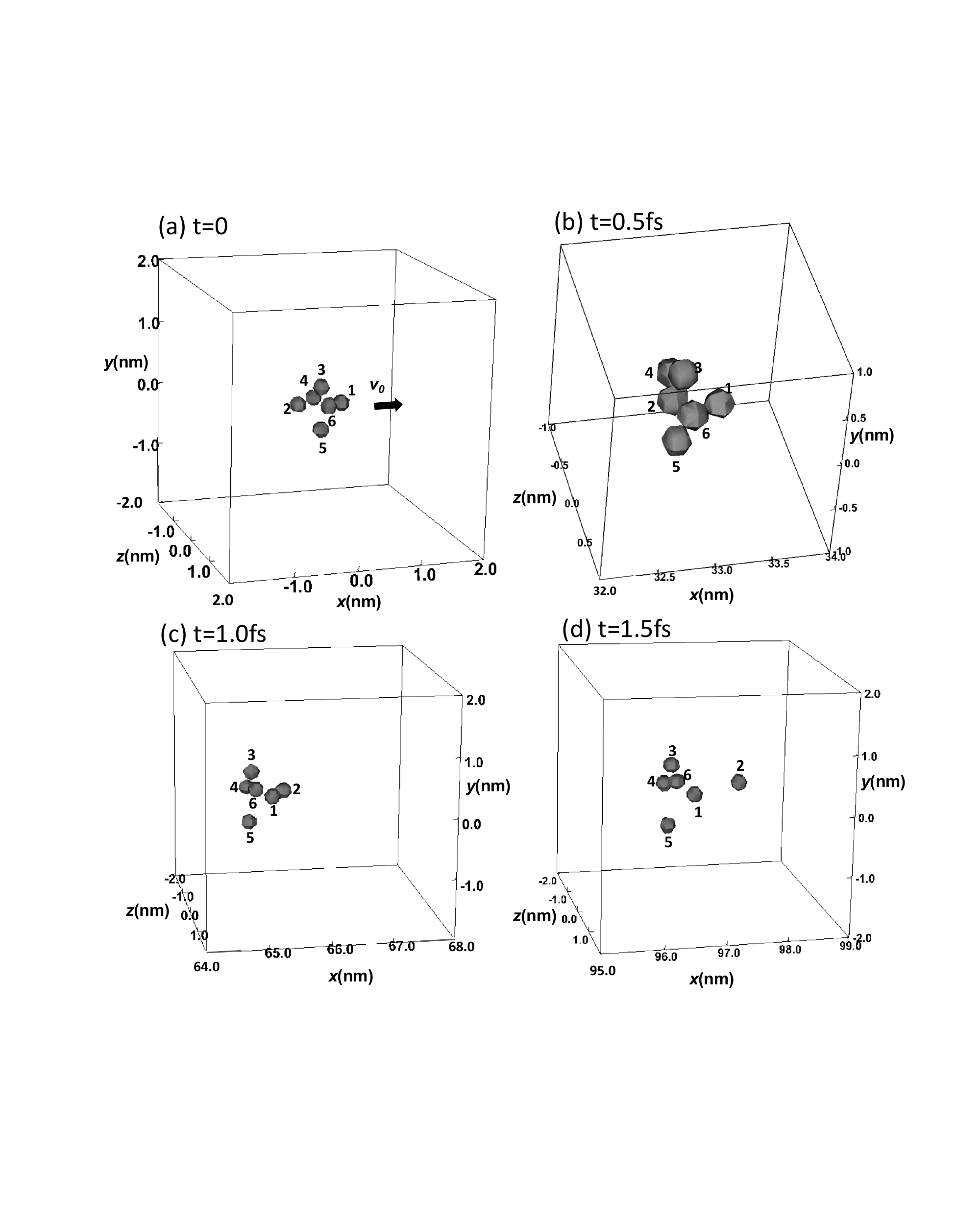}%
 \caption{A Si cluster, composed of 6 Si ions, interacts with a solid Al. The Si cluster speed is $v_{x}=0.221c$, and the distance between two Si ions is 5\r{A}. Figures show the 6 Si ion positions at (a) t=0, (b) 0.5fs, (c) 1.0fs and (d) 1.4fs.   \label{fig:Fig3}}
\end{figure}

\begin{figure}
 \includegraphics[width=90mm, angle=0]{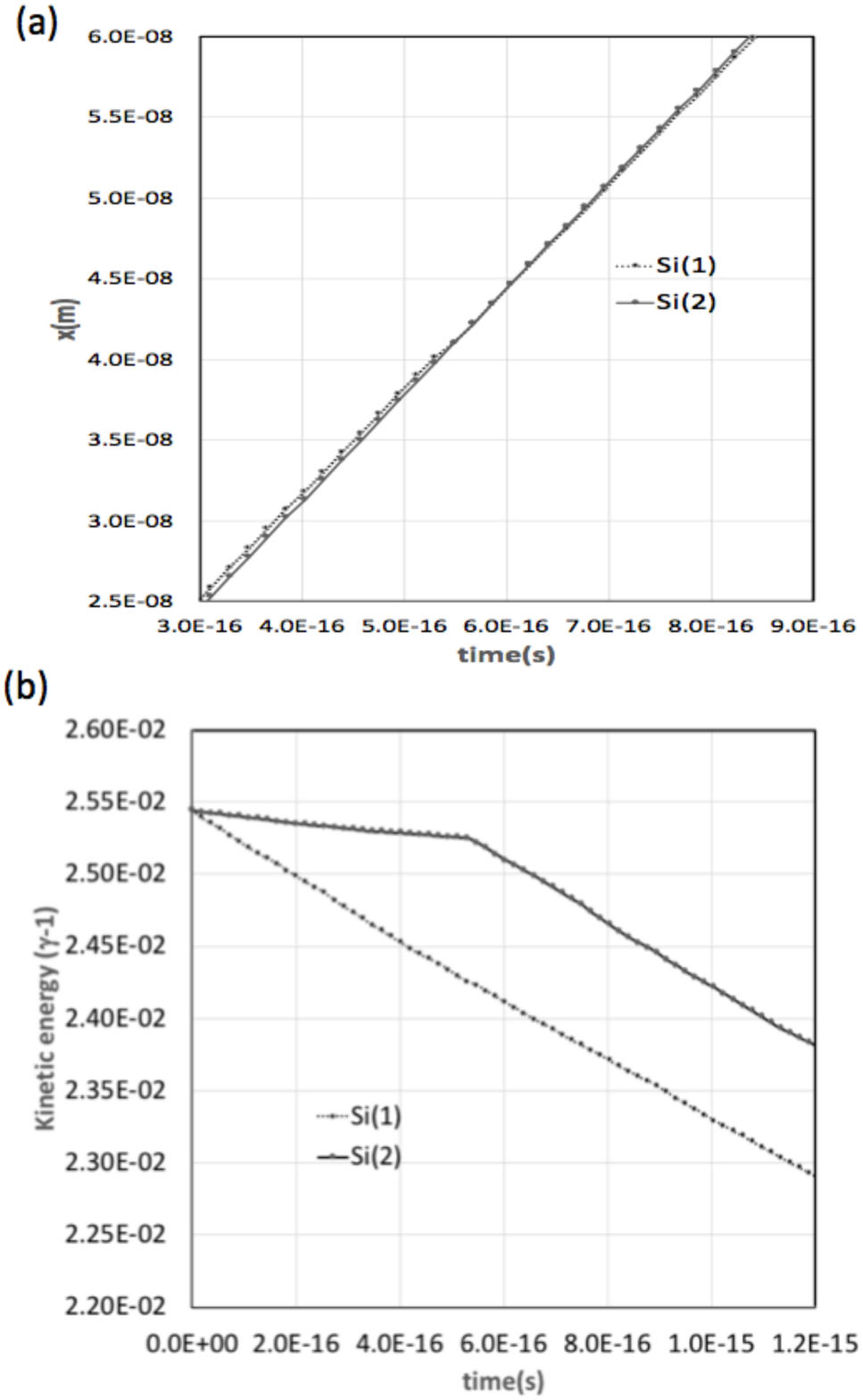}%
 \caption{(a) The history of the position in $x$, and (b) the kinetic energy history for Si(1) and Si(2) in one Si cluster, which is composed of 6 Si ions moving in the solid Al in the $x$ direction. The forward moving Si(1) dissipate its energy in the Al target, and the following Si(2) feels the acceleration wake field, generated by the forward moving Si(1).    \label{fig:Fig4}}
\end{figure}

However, the tail of the wake field is rather long as shown in Fig. \ref{fig:Fig2}(b). When two Si ions consisting a small cluster move together in parallel to the moving direction, the wake field has an influence to the other ion behind. Especially in our parameter range of the solid Al density and of the Si high speed, the wake field can cover Si ions just behind the forward-moving Si involved.  Figures \ref{fig:Fig3} present the time evolution of 6 Si ions in one Si cluster at  (a) t=0, (b) 0.5fs, (c) 1.0fs and (d) 1.4fs. The Si cluster composed of the 6 Si ions interacts with the solid Al. The Si cluster speed is $v_{x}=0.221c$, and the distance between two adjacent Si ions is 5\r{A}. In each figure, the number besides each Si ion shows the identity number of each Si ion. As we presented above, the forward moving Si(1), in which the number "(1)" shows the identity number for each Si, is caught and overtaken by the Si(2), which is located just behind of the Si(1). Each Si ion creates the wake field, and the electric field is localized in an elongated limited area as shown in Fig. \ref{fig:Fig2}. Among the 6 Si ions consisting of the Si cluster,  only the Si(2) feels the wake field by Si(1). The other 5 Si ions dissipate their energy as shown clearly in Fig. \ref{fig:Fig2}(d) and also in Fig. \ref{fig:Fig4}(b). Figure \ref{fig:Fig4} (a) presents the time sequence of the $x$ position  and Fig. \ref{fig:Fig4}(b) shows the history of the Si ion kinetic energy ($\gamma-1$) for Si(1) and Si(2). Each Si ion in one Si cluster creates the wake field, which is localized, and the kinetic energy of Si(1) loses its energy normally as shown in  Fig. \ref{fig:Fig4}(b). However, Si(2) is located just behind Si(1) (see Figs. \ref{fig:Fig3}), and feels the wake field by Si(1). The wake field by Si(2) is an acceleration field for Si(2), which is also creates the wake field behind Si(2). The acceleration and deceleration fields are almost balanced on Si(2), though the deceleration field is slightly strong so that Si(2) loses its energy slightly before the overtaking time of around $t$=0.57fs in this example (see Fig. \ref{fig:Fig4}(a)).  After the overtaking time, Si(2) starts to lose its energy normally, like Si(1) (see Fig. \ref{fig:Fig4}(b)).

The Si ion peculiar behavior discussed in this paper takes in a short time scale $\tau$ of about a few fs or less. In the present case, the collisional frequency $\nu_{ie}$ between a Si ion moving with 0.221$c$ and the electrons of the solid Al is much smaller than $1/\tau$. Therefore, the collective behavior of the high density Al electrons, that is, the wake field, mainly contributes to the Si cluster interaction in the solid Al during $\tau$. In addition, it should be pointed out that the overtaking behavior of the Si ions focused in this paper is difficult to observe experimentally, because the phenomenon is fast, and it is hard to distinguish one Si from another in one Si cluster. In addition, in the solid Al, one Si cluster alignment control is also difficult. The wake field is localized in space, and therefore, if one Si cluster rotates in its moving direction, Si(2) may not be covered by the Si(1) wake field. In this case, the Si ion overtaking behavior may not be taken place. 

In this paper we have presented the peculiar ion motion in a cluster interacting with the solid Al. In the context of the ion beam inertial fusion, the target electron density is so high, and the ion or cluster speed is very high: ${v_{0}}/{\sqrt{T_{e}/m_{e}}}\gg1$. Therefore, the wake field created by each ion in one cluster influences ions just behind the ion concerned, though the wake field strength is high because of the ion high speed. In this extreme situation, the ion, just behind the forward-moving ion in the cluster, catches up the forward ion, and moves ahead of the other forward ion. though the front ion looses its energy normally. When the target becomes hot and dilute, the perfect vicinage effect, that is, the stopping power enhancement would be expected.

\begin{acknowledgments}
We thank K. Takayama, K. Horioka, and J. Hasegawa for their discussions. This work was partly supported by Ministry of Education, Culture, Sports, Science and Technology (MEXT), Japan / U. S. Cooperation in Fusion Research and Development, and Center for Optical Research \& Education (CORE) in Utsunomiya University. 
\end{acknowledgments}

\nocite{*}
\bibliography{./References}

\providecommand{\noopsort}[1]{}\providecommand{\singleletter}[1]{#1}%
\begin{thebibliography}{29}%
\makeatletter
\providecommand \@ifxundefined [1]{%
 \@ifx{#1\undefined}
}%
\providecommand \@ifnum [1]{%
 \ifnum #1\expandafter \@firstoftwo
 \else \expandafter \@secondoftwo
 \fi
}%
\providecommand \@ifx [1]{%
 \ifx #1\expandafter \@firstoftwo
 \else \expandafter \@secondoftwo
 \fi
}%
\providecommand \natexlab [1]{#1}%
\providecommand \enquote  [1]{``#1''}%
\providecommand \bibnamefont  [1]{#1}%
\providecommand \bibfnamefont [1]{#1}%
\providecommand \citenamefont [1]{#1}%
\providecommand \href@noop [0]{\@secondoftwo}%
\providecommand \href [0]{\begingroup \@sanitize@url \@href}%
\providecommand \@href[1]{\@@startlink{#1}\@@href}%
\providecommand \@@href[1]{\endgroup#1\@@endlink}%
\providecommand \@sanitize@url [0]{\catcode `\\12\catcode `\$12\catcode
  `\&12\catcode `\#12\catcode `\^12\catcode `\_12\catcode `\%12\relax}%
\providecommand \@@startlink[1]{}%
\providecommand \@@endlink[0]{}%
\providecommand \url  [0]{\begingroup\@sanitize@url \@url }%
\providecommand \@url [1]{\endgroup\@href {#1}{\urlprefix }}%
\providecommand \urlprefix  [0]{URL }%
\providecommand \Eprint [0]{\href }%
\providecommand \doibase [0]{http://dx.doi.org/}%
\providecommand \selectlanguage [0]{\@gobble}%
\providecommand \bibinfo  [0]{\@secondoftwo}%
\providecommand \bibfield  [0]{\@secondoftwo}%
\providecommand \translation [1]{[#1]}%
\providecommand \BibitemOpen [0]{}%
\providecommand \bibitemStop [0]{}%
\providecommand \bibitemNoStop [0]{.\EOS\space}%
\providecommand \EOS [0]{\spacefactor3000\relax}%
\providecommand \BibitemShut  [1]{\csname bibitem#1\endcsname}%
\let\auto@bib@innerbib\@empty
\bibitem [{\citenamefont {Wang}\ \emph {et~al.}(2000)\citenamefont {Wang},
  \citenamefont {Qiu},\ and\ \citenamefont {Mi\u{s}kovi\'{c}}}]{Wang2000}%
  \BibitemOpen
  \bibfield  {author} {\bibinfo {author} {\bibfnamefont {Y.-N.}\ \bibnamefont
  {Wang}}, \bibinfo {author} {\bibfnamefont {H.-T.}\ \bibnamefont {Qiu}}, \
  and\ \bibinfo {author} {\bibfnamefont {Z.~L.}\ \bibnamefont
  {Mi\u{s}kovi\'{c}}},\ }\href@noop {} {\bibfield  {journal} {\bibinfo
  {journal} {Phys. Rev. Lett.}\ }\textbf {\bibinfo {volume} {85}},\ \bibinfo
  {pages} {1448} (\bibinfo {year} {2000})}\BibitemShut {NoStop}%
\bibitem [{\citenamefont {Wang}\ \emph {et~al.}(2018)\citenamefont {Wang} \emph
  {et~al.}}]{Wang2018}%
  \BibitemOpen
  \bibfield  {author} {\bibinfo {author} {\bibfnamefont {G.}~\bibnamefont
  {Wang}} \emph {et~al.},\ }\href@noop {} {\bibfield  {journal} {\bibinfo
  {journal} {Matter and Radiation at Extreme}\ }\textbf {\bibinfo {volume}
  {3}},\ \bibinfo {pages} {67} (\bibinfo {year} {2018})}\BibitemShut {NoStop}%
\bibitem [{\citenamefont {Ditmire}\ \emph {et~al.}(1997)\citenamefont {Ditmire}
  \emph {et~al.}}]{Ditmire1}%
  \BibitemOpen
  \bibfield  {author} {\bibinfo {author} {\bibfnamefont {T.}~\bibnamefont
  {Ditmire}} \emph {et~al.},\ }\href@noop {} {\bibfield  {journal} {\bibinfo
  {journal} {Nature}\ }\textbf {\bibinfo {volume} {386}},\ \bibinfo {pages}
  {54} (\bibinfo {year} {1997})}\BibitemShut {NoStop}%
\bibitem [{\citenamefont {Ditmire}\ \emph {et~al.}(1999)\citenamefont {Ditmire}
  \emph {et~al.}}]{Ditmire2}%
  \BibitemOpen
  \bibfield  {author} {\bibinfo {author} {\bibfnamefont {T.}~\bibnamefont
  {Ditmire}} \emph {et~al.},\ }\href@noop {} {\bibfield  {journal} {\bibinfo
  {journal} {Nature}\ }\textbf {\bibinfo {volume} {398}},\ \bibinfo {pages}
  {489} (\bibinfo {year} {1999})}\BibitemShut {NoStop}%
\bibitem [{\citenamefont {Brunelle}\ \emph {et~al.}(1997)\citenamefont
  {Brunelle} \emph {et~al.}}]{Burunelle1}%
  \BibitemOpen
  \bibfield  {author} {\bibinfo {author} {\bibfnamefont {A.}~\bibnamefont
  {Brunelle}} \emph {et~al.},\ }\href@noop {} {\bibfield  {journal} {\bibinfo
  {journal} {Nucl. Instrum. Methofs Phys. Res. B}\ }\textbf {\bibinfo {volume}
  {125}},\ \bibinfo {pages} {207} (\bibinfo {year} {1997})}\BibitemShut
  {NoStop}%
\bibitem [{\citenamefont {Takayama}\ \emph {et~al.}(2005)\citenamefont
  {Takayama} \emph {et~al.}}]{Takayama1}%
  \BibitemOpen
  \bibfield  {author} {\bibinfo {author} {\bibfnamefont {K.}~\bibnamefont
  {Takayama}} \emph {et~al.},\ }\href@noop {} {\bibfield  {journal} {\bibinfo
  {journal} {Phys. Rev. Lett.}\ }\textbf {\bibinfo {volume} {94}},\ \bibinfo
  {pages} {144801} (\bibinfo {year} {2005})}\BibitemShut {NoStop}%
\bibitem [{\citenamefont {Kawata}\ \emph {et~al.}(2016)\citenamefont {Kawata},
  \citenamefont {Karino},\ and\ \citenamefont {Ogoyski}}]{Kawata1}%
  \BibitemOpen
  \bibfield  {author} {\bibinfo {author} {\bibfnamefont {S.}~\bibnamefont
  {Kawata}}, \bibinfo {author} {\bibfnamefont {T.}~\bibnamefont {Karino}}, \
  and\ \bibinfo {author} {\bibfnamefont {A.~I.}\ \bibnamefont {Ogoyski}},\
  }\href@noop {} {\bibfield  {journal} {\bibinfo  {journal} {Matter and
  Radiation at Extreme}\ }\textbf {\bibinfo {volume} {1}},\ \bibinfo {pages}
  {89} (\bibinfo {year} {2016})}\BibitemShut {NoStop}%
\bibitem [{\citenamefont {Clauser}(1975)}]{Clauser1}%
  \BibitemOpen
  \bibfield  {author} {\bibinfo {author} {\bibfnamefont {M.~J.}\ \bibnamefont
  {Clauser}},\ }\href@noop {} {\bibfield  {journal} {\bibinfo  {journal} {Phys.
  Rev. Lett.}\ }\textbf {\bibinfo {volume} {35}},\ \bibinfo {pages} {848}
  (\bibinfo {year} {1975})}\BibitemShut {NoStop}%
\bibitem [{\citenamefont {Hofmann}(2018)}]{IHofmann1}%
  \BibitemOpen
  \bibfield  {author} {\bibinfo {author} {\bibfnamefont {I.}~\bibnamefont
  {Hofmann}},\ }\href@noop {} {\bibfield  {journal} {\bibinfo  {journal}
  {Matter and Radiation at Extreme}\ }\textbf {\bibinfo {volume} {3}},\
  \bibinfo {pages} {1} (\bibinfo {year} {2018})}\BibitemShut {NoStop}%
\bibitem [{\citenamefont {Tahir}\ \emph {et~al.}(1996)\citenamefont {Tahir}
  \emph {et~al.}}]{Tahir1}%
  \BibitemOpen
  \bibfield  {author} {\bibinfo {author} {\bibfnamefont {N.~A.}\ \bibnamefont
  {Tahir}} \emph {et~al.},\ }\href@noop {} {\bibfield  {journal} {\bibinfo
  {journal} {Phys. Plasmas}\ }\textbf {\bibinfo {volume} {4}},\ \bibinfo
  {pages} {796} (\bibinfo {year} {1996})}\BibitemShut {NoStop}%
\bibitem [{\citenamefont {Deutsch}\ \emph {et~al.}(1997)\citenamefont {Deutsch}
  \emph {et~al.}}]{Deutsch1}%
  \BibitemOpen
  \bibfield  {author} {\bibinfo {author} {\bibfnamefont {C.}~\bibnamefont
  {Deutsch}} \emph {et~al.},\ }\href@noop {} {\bibfield  {journal} {\bibinfo
  {journal} {Fusion Tech.}\ }\textbf {\bibinfo {volume} {31}},\ \bibinfo
  {pages} {1} (\bibinfo {year} {1997})}\BibitemShut {NoStop}%
\bibitem [{\citenamefont {Zwicknagel}\ and\ \citenamefont
  {Deutsch}(1997)}]{Zwicknagel1}%
  \BibitemOpen
  \bibfield  {author} {\bibinfo {author} {\bibfnamefont {G.}~\bibnamefont
  {Zwicknagel}}\ and\ \bibinfo {author} {\bibfnamefont {C.}~\bibnamefont
  {Deutsch}},\ }\href@noop {} {\bibfield  {journal} {\bibinfo  {journal} {Phys.
  Rev. E}\ }\textbf {\bibinfo {volume} {56}},\ \bibinfo {pages} {970} (\bibinfo
  {year} {1997})}\BibitemShut {NoStop}%
\bibitem [{\citenamefont {Bret}\ and\ \citenamefont {Deutsch}(2008)}]{Bret1}%
  \BibitemOpen
  \bibfield  {author} {\bibinfo {author} {\bibfnamefont {A.}~\bibnamefont
  {Bret}}\ and\ \bibinfo {author} {\bibfnamefont {C.}~\bibnamefont {Deutsch}},\
  }\href@noop {} {\bibfield  {journal} {\bibinfo  {journal} {J. Plasma Phys.}\
  }\textbf {\bibinfo {volume} {74}},\ \bibinfo {pages} {595} (\bibinfo {year}
  {2008})}\BibitemShut {NoStop}%
\bibitem [{\citenamefont {Nardi}\ \emph {et~al.}(2002)\citenamefont {Nardi}
  \emph {et~al.}}]{Nardi1}%
  \BibitemOpen
  \bibfield  {author} {\bibinfo {author} {\bibfnamefont {E.}~\bibnamefont
  {Nardi}} \emph {et~al.},\ }\href@noop {} {\bibfield  {journal} {\bibinfo
  {journal} {Phys. Rev. A}\ }\textbf {\bibinfo {volume} {66}},\ \bibinfo
  {pages} {013201} (\bibinfo {year} {2002})}\BibitemShut {NoStop}%
\bibitem [{\citenamefont {Nardi}\ and\ \citenamefont {Zinamon}(1995)}]{Nardi2}%
  \BibitemOpen
  \bibfield  {author} {\bibinfo {author} {\bibfnamefont {E.}~\bibnamefont
  {Nardi}}\ and\ \bibinfo {author} {\bibfnamefont {Z.}~\bibnamefont
  {Zinamon}},\ }\href@noop {} {\bibfield  {journal} {\bibinfo  {journal} {Phys.
  Rev. A}\ }\textbf {\bibinfo {volume} {51}},\ \bibinfo {pages} {R3407}
  (\bibinfo {year} {1995})}\BibitemShut {NoStop}%
\bibitem [{\citenamefont {Ben-Hamu}\ \emph {et~al.}(1997)\citenamefont
  {Ben-Hamu} \emph {et~al.}}]{Ben-Hamu1}%
  \BibitemOpen
  \bibfield  {author} {\bibinfo {author} {\bibfnamefont {D.}~\bibnamefont
  {Ben-Hamu}} \emph {et~al.},\ }\href@noop {} {\bibfield  {journal} {\bibinfo
  {journal} {Phys. Rev. A}\ }\textbf {\bibinfo {volume} {56}},\ \bibinfo
  {pages} {4786} (\bibinfo {year} {1997})}\BibitemShut {NoStop}%
\bibitem [{\citenamefont {Arista}(2000)}]{Arista1}%
  \BibitemOpen
  \bibfield  {author} {\bibinfo {author} {\bibfnamefont {N.~R.}\ \bibnamefont
  {Arista}},\ }\href@noop {} {\bibfield  {journal} {\bibinfo  {journal} {Nucl.
  Instrum. Methods Phys. Res. B}\ }\textbf {\bibinfo {volume} {164-165}},\
  \bibinfo {pages} {108} (\bibinfo {year} {2000})}\BibitemShut {NoStop}%
\bibitem [{\citenamefont {Brandt}\ \emph {et~al.}(1974)\citenamefont {Brandt},
  \citenamefont {Ratkowski},\ and\ \citenamefont {Richie}}]{Brandt1}%
  \BibitemOpen
  \bibfield  {author} {\bibinfo {author} {\bibfnamefont {W.}~\bibnamefont
  {Brandt}}, \bibinfo {author} {\bibfnamefont {A.}~\bibnamefont {Ratkowski}}, \
  and\ \bibinfo {author} {\bibfnamefont {R.~H.}\ \bibnamefont {Richie}},\
  }\href@noop {} {\bibfield  {journal} {\bibinfo  {journal} {Phys. Rev. Lett.}\
  }\textbf {\bibinfo {volume} {33}},\ \bibinfo {pages} {1325} (\bibinfo {year}
  {1974})}\BibitemShut {NoStop}%
\bibitem [{\citenamefont {Horioka}(2018)}]{Horioka1}%
  \BibitemOpen
  \bibfield  {author} {\bibinfo {author} {\bibfnamefont {K.}~\bibnamefont
  {Horioka}},\ }\href@noop {} {\bibfield  {journal} {\bibinfo  {journal}
  {Matter and Radiation at Extreme}\ }\textbf {\bibinfo {volume} {3}},\
  \bibinfo {pages} {12} (\bibinfo {year} {2018})}\BibitemShut {NoStop}%
\bibitem [{\citenamefont {Chenevier}\ \emph {et~al.}(1973)\citenamefont
  {Chenevier}, \citenamefont {Dolique},\ and\ \citenamefont
  {Per\`{e}s}}]{Chenevier1}%
  \BibitemOpen
  \bibfield  {author} {\bibinfo {author} {\bibfnamefont {P.}~\bibnamefont
  {Chenevier}}, \bibinfo {author} {\bibfnamefont {J.~M.}\ \bibnamefont
  {Dolique}}, \ and\ \bibinfo {author} {\bibfnamefont {H.}~\bibnamefont
  {Per\`{e}s}},\ }\href@noop {} {\bibfield  {journal} {\bibinfo  {journal} {J.
  Plasma Phys.}\ }\textbf {\bibinfo {volume} {10}},\ \bibinfo {pages} {185}
  (\bibinfo {year} {1973})}\BibitemShut {NoStop}%
\bibitem [{\citenamefont {Wang}\ \emph {et~al.}(1981)\citenamefont {Wang},
  \citenamefont {Joyce},\ and\ \citenamefont {Nicholson}}]{CLWang1}%
  \BibitemOpen
  \bibfield  {author} {\bibinfo {author} {\bibfnamefont {C.-L.}\ \bibnamefont
  {Wang}}, \bibinfo {author} {\bibfnamefont {G.}~\bibnamefont {Joyce}}, \ and\
  \bibinfo {author} {\bibfnamefont {D.~R.}\ \bibnamefont {Nicholson}},\
  }\href@noop {} {\bibfield  {journal} {\bibinfo  {journal} {J. Plasma Phys.}\
  }\textbf {\bibinfo {volume} {25}},\ \bibinfo {pages} {225} (\bibinfo {year}
  {1981})}\BibitemShut {NoStop}%
\bibitem [{\citenamefont {Echenique}\ \emph {et~al.}(1990)\citenamefont
  {Echenique}, \citenamefont {Flores},\ and\ \citenamefont
  {Ritche}}]{Echenique1}%
  \BibitemOpen
  \bibfield  {author} {\bibinfo {author} {\bibfnamefont {P.~M.}\ \bibnamefont
  {Echenique}}, \bibinfo {author} {\bibfnamefont {F.}~\bibnamefont {Flores}}, \
  and\ \bibinfo {author} {\bibfnamefont {R.~H.}\ \bibnamefont {Ritche}},\
  }\href@noop {} {\bibfield  {journal} {\bibinfo  {journal} {Solid State
  Phys.}\ }\textbf {\bibinfo {volume} {43}},\ \bibinfo {pages} {229} (\bibinfo
  {year} {1990})}\BibitemShut {NoStop}%
\bibitem [{\citenamefont {Ichimaru}(2004)}]{Ichimaru1}%
  \BibitemOpen
  \bibfield  {author} {\bibinfo {author} {\bibfnamefont {S.}~\bibnamefont
  {Ichimaru}},\ }\href@noop {} {\emph {\bibinfo {title} {Statistical Plasma
  Physics: Basic Principles}}},\ \bibinfo {edition} {1st}\ ed.,\ \bibinfo
  {series} {Frontiers in Physics}, Vol.~\bibinfo {volume} {1}\ (\bibinfo
  {publisher} {Westview Press},\ \bibinfo {address} {Colorado},\ \bibinfo
  {year} {2004})\BibitemShut {NoStop}%
\bibitem [{\citenamefont {Arista}\ and\ \citenamefont
  {Brabdt}(1984)}]{Arista2}%
  \BibitemOpen
  \bibfield  {author} {\bibinfo {author} {\bibfnamefont {N.~R.}\ \bibnamefont
  {Arista}}\ and\ \bibinfo {author} {\bibfnamefont {W.}~\bibnamefont
  {Brabdt}},\ }\href@noop {} {\bibfield  {journal} {\bibinfo  {journal} {Phys.
  Rev. A}\ }\textbf {\bibinfo {volume} {29}},\ \bibinfo {pages} {1471}
  (\bibinfo {year} {1984})}\BibitemShut {NoStop}%
\bibitem [{\citenamefont {Maynard}\ and\ \citenamefont
  {Deutsch}(1982)}]{Maynard1}%
  \BibitemOpen
  \bibfield  {author} {\bibinfo {author} {\bibfnamefont {G.}~\bibnamefont
  {Maynard}}\ and\ \bibinfo {author} {\bibfnamefont {C.}~\bibnamefont
  {Deutsch}},\ }\href@noop {} {\bibfield  {journal} {\bibinfo  {journal} {Phys.
  Rev. A}\ }\textbf {\bibinfo {volume} {26}},\ \bibinfo {pages} {665} (\bibinfo
  {year} {1982})}\BibitemShut {NoStop}%
\bibitem [{\citenamefont {Arista}\ and\ \citenamefont
  {Brabdt}(1981)}]{Arista3}%
  \BibitemOpen
  \bibfield  {author} {\bibinfo {author} {\bibfnamefont {N.~R.}\ \bibnamefont
  {Arista}}\ and\ \bibinfo {author} {\bibfnamefont {W.}~\bibnamefont
  {Brabdt}},\ }\href@noop {} {\bibfield  {journal} {\bibinfo  {journal} {Phys.
  Rev. A}\ }\textbf {\bibinfo {volume} {23}},\ \bibinfo {pages} {1898}
  (\bibinfo {year} {1981})}\BibitemShut {NoStop}%
\bibitem [{\citenamefont {Kittel}(2004)}]{Kittel1}%
  \BibitemOpen
  \bibfield  {author} {\bibinfo {author} {\bibfnamefont {C.}~\bibnamefont
  {Kittel}},\ }\href@noop {} {\emph {\bibinfo {title} {Introduction to Solid
  State Physics}}},\ \bibinfo {edition} {8th}\ ed.\ (\bibinfo  {publisher}
  {John Wiley and Sons},\ \bibinfo {address} {New Jersey},\ \bibinfo {year}
  {2004})\BibitemShut {NoStop}%
\bibitem [{\citenamefont {Ridgers}\ \emph {et~al.}(2014)\citenamefont {Ridgers}
  \emph {et~al.}}]{EPOCH1}%
  \BibitemOpen
  \bibfield  {author} {\bibinfo {author} {\bibfnamefont {C.}~\bibnamefont
  {Ridgers}} \emph {et~al.},\ }\href@noop {} {\bibfield  {journal} {\bibinfo
  {journal} {J. Comput. Phys.}\ }\textbf {\bibinfo {volume} {260}},\ \bibinfo
  {pages} {273} (\bibinfo {year} {2014})}\BibitemShut {NoStop}%
\bibitem [{\citenamefont {Arber}\ \emph {et~al.}(2015)\citenamefont {Arber}
  \emph {et~al.}}]{EPOCH2}%
  \BibitemOpen
  \bibfield  {author} {\bibinfo {author} {\bibfnamefont {T.~D.}\ \bibnamefont
  {Arber}} \emph {et~al.},\ }\href@noop {} {\bibfield  {journal} {\bibinfo
  {journal} {Plasma Phys. Control. Fusion}\ }\textbf {\bibinfo {volume} {57}},\
  \bibinfo {pages} {113001} (\bibinfo {year} {2015})}\BibitemShut {NoStop}%
\end{thebibliography}%

\end{document}